\documentclass[preprint,showpacs,preprintnumbers,amsmath,amssymb]{revtex4}

\newcommand\rf[1]{(\ref{eq:#1})}
\newcommand\lab[1]{\label{eq:#1}}

\newcommand\br{\begin{eqnarray}}
\newcommand\er{\end{eqnarray}}
\newcommand\be{\begin{equation}}
\newcommand\ee{\end{equation}}

\newcommand\lb{\lbrack}
\newcommand\rb{\rbrack}

\newcommand\bc{\begin{center}}
\newcommand\ec{\end{center}}

\newcommand\pa{\partial}

\newcommand{\ct}[1]{\cite{#1}}
\newcommand{\bib}[1]{\bibitem{#1}}
\newcommand\PRL[3]{\textsl{Phys. Rev. Lett.} \textbf{#1}, #3 (#2)}

\begin{document}

\preprint{arxiv:[hep-th]}

\title{Continuous Axion Photon Duality and its Consequences }

\author{E.I. Guendelman}%
\email{guendel@bgumail.bgu.ac.il}
\affiliation{%
Department of Physics, Ben-Gurion University of the Negev \\
P.O.Box 653, IL-84105 ~Beer-Sheva, Israel
}%

\begin{abstract}
The axion photon system in an external magnetic field, when for example considered with the 
geometry of the experiments exploring axion photon mixing, displays a continuous axion-photon 
duality symmetry in the limit the axion mass is neglected. The conservation law that follows 
from this symmetry is obtained. The magnetic field interaction is seen to be equivalent to first order
to the interaction of a complex charged field with an external electric potential, where this 
ficticious "electric potential" is proportional to the external magnetic field. This allows to  solve
for the scattering amplitudes using already known scalar QED results. It 
is argued that in more generic conditions (not just related to these experiments) axion-photon
condensation could be obtained for high magnetic fields. Finally an exact constraint originating from
the current conservation on the amplitudes of reflected and transmited waves is obtained.
\end{abstract}


\maketitle

\section{Introduction}

The axion \ct{weinberg}  was introduced in order to solve the strong CP problem.
Since then the axion has been postulated as a candidate for the dark matter also.
A great number of ideas and experiments for the search this particle have been proposed
\ct{Goldman}.

One particular feature of the axion field $\phi$ is its coupling to the photon 
through an interaction term of the form 
$g \phi \epsilon^{\mu \nu \alpha \beta}F_{\mu \nu} F_{\alpha \beta}$.
In fact a coupling of this sort is natural for any pseudoscalar interacting with
electromagnetism , as is the case of the neutral pion coupling to photons (which
as a consequence of this interaction decays into two photons).

A way to explore for observable consequences of the coupling of a light scalar 
to the photon in this way is to subject a beam of photons to a very strong magnetic field.

This affects the optical properties of light which could lead to testable consequences\ct{PVLAS}. 
Also, the produced axions could be
responsible for the "light shining through a wall phenomena ", which are is obtained by first
producing axions out of photons in a strong magnetic field region, then subjecting the mixed beam of photons
and axions to an absorbing wall for photons, but almost totally transparent to axions due to their weak 
interacting properties which can then
go through behind this "wall", applying then another magnetic field one can recover once again some photons 
from the produced axions \ct{LSTW}.

\section{Action and Equations of Motion}
The action principle describing the relevant light pseudoscalar coupling to the photon is
\be
S =  \int d^{4}x 
\Bigl\lb -\frac{1}{4}F^{\mu\nu}F_{\mu\nu} + \frac{1}{2}\pa_{\mu}\phi \pa^{\mu}\phi - 
\frac{1}{2}m^{2}\phi^{2} + 
\frac{g}{2}g \phi \epsilon^{\mu \nu \alpha \beta}F_{\mu \nu} F_{\alpha \beta}\Bigr\rb
\lab{axion photon ac}
\ee

We now specialize to the case where we consider an electromagnetic field with propagation only along the z-direction
and where a strong magnetic field pointing in the x-direction is present. This field may have an arbitrary space dependence in z, but it is assumed to be time independent. In the case the magnetic field is constant, see for example \ct{Ansoldi} for general solutions.

For the small perturbations we consider only small quadratic terms in the action for the axion fields and the electromagnetic field, following the method of for example Ref. \ct{Ansoldi}, but now considering a static magnetic field pointing in the x direction  having
arbitrary z dependence and specializing to z dependent electromagnetic field perturbations and axion fields. This means that the interaction between the background field , the axion and photon fields
reduces to
 
\be
S_I =  \int d^{4}x 
\Bigl\lb \beta \phi E_x\Bigr\rb
\lab{axion photon int}
\ee

where $\beta = gB(z) $. Choosing the temporal gauge for the photon excitations and considering only the x-polarization for the electromagnetic waves, since only this polarization couples to the axion, we get the following 1+1 effective dimensional action
(A being the x-polarization of the photon)

\be
S_2 =  \int dzdt 
\Bigl\lb  \frac{1}{2}\pa_{\mu}A \pa^{\mu}A+ \frac{1}{2}\pa_{\mu}\phi \pa^{\mu}\phi - 
\frac{1}{2}m^{2}\phi^{2} + \beta \phi \pa_{t} A
\Bigr\rb
\lab{2 action}
\ee

($A=A(t,z)$, $\phi =\phi(t,z)$), which leads to the equations

\be
\pa_{\mu}\pa^{\mu}\phi + m^{2}\phi =  \beta \pa_{t} A
\lab{eq. ax}
\ee

\be
\pa_{\mu} \pa^{\mu}A = - \beta \pa_{t}\phi 
\lab{eq. photon}
\ee

As it is known, in temporal gauge, the action principle cannot reproduce the Gauss 
constraint (here with a charge density obtained from the axion photon coupling) and has
to be impossed as a complementary condition. However this 
constraint is automatically satisfied here just because of the type of dynamical reduction
employed and does not need to be considered  anymore.

\section{The Continuous Axion Photon Duality Symmetry and the Scalar QED analogy}
Without assuming any particular z-dependence for $\beta$, but still insisting that 
it will be static, we see that in the case $m=0$, we discover a continuous axion 
photon duality symmetry, since,

1. The kinetic terms of the photon and
axion allow for a rotational $O(2)$ symmetry in the axion-photon field space.

2. The interaction term, after dropping  a total time derivative can also be expressed in 
an $O(2)$ symmetric way as follows

\be
S_I =  \frac{1}{2} \int dzdt 
\beta \Bigl\lb \phi \pa_{t} A - A \pa_{t}\phi \Bigr\rb
\lab{axion photon int2}
\ee

The axion photon symmetry is in the infinitesimal limit
\be
\delta A = \epsilon \phi, \delta \phi = - \epsilon A
\lab{axion photon symm}
\ee

where $\epsilon$ is a small number. Using Noether`s theorem, this leads to the 
conserved current $j_{\mu}$, with components given by

\be
j_{0} = A \pa_{t}\phi - \phi \pa_{t} A + \frac{\beta}{2}(A^{2} + \phi^{2} )
\lab{axion photon density}
\ee
and 
\be
j_{i} = A \pa_{i}\phi - \phi \pa_{i} A 
\lab{axion photon current}
\ee
defining the complex field $\psi$ as
\be
\psi = \frac{1}{\sqrt{2}}(\phi + iA)
\lab{axion photon complex}
\ee
we see that 
in terms of this complex field, the axion photon density takes the form
\be
j_{0} = i( \psi^{*}\pa_{t}\psi - \psi \pa_{t} \psi^{*}) +  \beta \psi^{*}\psi 
\lab{axion photon density complex}
\ee

We observe that to first order in $\beta$, \rf{axion photon int2} represents the
interaction of the magnetic field with the "axion photon density" \rf{axion photon density}, \rf{axion photon density complex} 
and also this interaction has the same form as that of scalar QED with an external "electric " field to first order.
In fact the magnetic field or more precisely $\beta /2$ appears to play the role of external electric potential 
that couples to the axion photon density \rf{axion photon density},\rf{axion photon density complex} which appears then to play the role
of an electric charge density . From this analogy one can obtain without effort the scattering amplitudes,
just using the known results from the scattering of charged scalar particles under the influence of an external
static electric potential, see for example \ct{Bjorken-Drell} where an external one dimensional square electric potential 
is studied and also the perturbative scattering treatment for arbitrary shape potential is also given.
Any study of a one dimensional electric potential for the charged scalar particles translates then into a corresponding
solution for the axion photon system. For example one could use the results from the "cusp potential" \ct{Villalba}(that has as 
a limit the delta function potential) in the case of charged scalars to get results for a similar shape magnetic field in the
axion photon system.

One should notice however that the natural initial states used for example in "light shininig through the
wall experiments" , like an initial photon and no axion involved, is not going to have a well defined axion photon charge in the second
quantized theory (although its average value appears zero), so the 
S matrix has to be presented in a different basis than that of normal QED . This is similar to the difference between working with linear polarizations as opposed to circular polarizations in ordinary optics, except that here we talk about polarizations in the axion photon space.
In fact pure axion and pure photon initial states correspond to symmetric and antisymmetric linear combinations of particle and antiparticle 
in the analog QED language. The reason these linear combinations are not going to be mantained in the presence on $B$  in the analog 
QED language, is that the analog external electric potential breaks the symmetry between particle and antiparticle and therefore will not
mantain in time the symmetric or antisymmetric combinations.  

One immediate consequence of \rf{axion photon density} is the existence of a non trivial
charge density in the case of the zero energy momentum solutions, in fact for constant $A$
and $\phi$ (which are solutions of the equations of motion if $m=0$), we get a non trivial 
charge density at any point having the value
$\frac{\beta}{2}(A^{2} + \phi^{2} )$. This is suggestive of the possibility that strong 
magnetic fields could be responsible for axion-photon condensations in more general settings
than the one explained here motivated by the axion-photon conversion experiments.

From the point of view of the axion-photon conversion experiments,
the symmetry \rf{axion photon symm} and its finite form , which is just
a rotation in the axion-photon space, implies a corresponding symmetry
of the axion-photon conversion amplitudes, for the limit $\omega >>m$.

In terms of the complex field, the axion photon current takes the form
\be
j_{k} = i( \psi^{*}\pa_{k}\psi - \psi \pa_{k} \psi^{*}) 
\lab{axion photon current complex}
\ee
Now, let us take any non trivial magnetic field dependence, but such that for $z<0$ and
for $z>L$, there is no magnetic field. Now consider  $\psi = exp(-i\omega t)\Psi(z)$, with
for example 
(taking the amplitude of the incident wave to have modulus 1),
\be
 \Psi(z) = exp(i\omega z+\varphi) + \Psi_{R}exp(-i\omega z)
\lab{reflected}
\ee
for $z<0$
and
\be
 \Psi(z) =  \Psi_{T}exp(i\omega z)
\lab{transmitted}
\ee
for $z>L$

Then in this stationary situation, from current conservation, we obtain that the current calculated for
$z<0$ must equal the current calculated for $z>L$, which leads to the constraint
\be
  (\Psi_{R})^{*}(\Psi_{R}) + (\Psi_{T})^{*}(\Psi_{T}) =1
\lab{constraint}
\ee

which represents a sort of "axion photon" unitarity constraint.
This type of field configuration is  given as an example only, which may  not be general
to all possible experiments. This is because in our formalism both real and imaginary parts 
have direct physical meaning and for example one should consider cases where the incident
$\psi$ is real, or pure imaginary, which are not addressed in the previous example.

\section{Conclusions}
The limit of zero axion mass when considering the scattering of axions and photons with the geometry relevant to the
axion-photon mixing experiments reveals a continuous axion photon duality symmetry.
This symmetry leads to a conserved current and then one observes that the interaction of the exteral magnetic field 
with the axion and photon is, to  first order in the magnetic field, of the form of the first order in coupling constant 
interaction of charged scalars with an external electric scalar potential. Here the role of this ficticious external 
electric scalar potential being played (up to a constant) by the external magnetic field. 

Pure axion and pure photon initial states correspond to symmetric and antisymmetric linear combinations of particle and antiparticle 
in the analog QED language. Notice in this respect that charge conjugation of \rf{axion photon complex} corresponds to sign reversal of the photon field. The reason these linear combinations are not going to be mantained in the presence on a non trivial $B$ in the analog 
QED language, is that the analog external electric potential breaks the symmetry between particle and antiparticle and therefore will not
mantain in time the symmetric or antisymmetric combinations.

It is possible that there will 
be situations, not related to axion photon mixing experiments, where at high magnetic fields, there will be some kind
of axion photon condensation.

Finally, we have seen an interesting constraint obtained fron the conservation on the
axion - photon current for a  particular case of a stationary wave incident on a region of magnetic field and leading to reflected 
and transmitted waves.

In a future publication we will study all these aspects in more details.

\section*{Acknowledgments}

I would like to thank Stephen Adler for a great number of exchanges and for clarifications from his part on diverse aspects of the subject 
of axion-photon mixing experiments.


\end{document}